\documentclass[12pt]{iopart}

\usepackage[dvips]{graphicx}

\begin{document}

\title{Comment on the Relativistic Galactic Model by Cooperstock and Tieu}

\author{Dylan Menzies$^1$ and Grant J. Mathews$^2$}

\address{$^{(1)}$ Department of Computer Science and Engineering, De Montfort University, Leicester UK \\
$^{(2)}$ Department of Physics, Center for Astrophysics, University of Notre Dame, Notre Dame, IN 46556}

\eads{\mailto{dylan@dmu.ac.uk}, \mailto{gmathews@nd.edu}}

\begin{abstract}
It has recently been suggested that observed galaxy rotation curves can be accounted for by general relativity without recourse to dark-matter halos. A number of objections have been raised, which have been addressed by the authors. Here, the calculation of tangential velocity is questioned.
\end{abstract}

\maketitle

\section{Galactic tangential velocity}

Cooperstock and Tieu have proposed a relativistic model of galaxies, \cite{CT}, which they claim can explain the observed flat rotation curves without the need for substantial additional unseen mass. They have refined the model, \cite{response2}, and responded, \cite{response, response2}, to a number of objections \cite{korz, VL, garf, cross}. 

Here we bring into question the derivation of the tangential velocity, $V(r)$, in the galaxy model. According to the authors,

\begin{equation}
V = r \omega~~,
\end{equation}
where $\omega$ defines the coordinate transformation,
\begin{equation}
\bar{\phi} = \phi + \omega(r,z) t~~,
\end{equation}
that locally diaganolizes the metric. In general it is easy to see this gives
\begin{equation}
\label{eq:omega}
\omega = \frac{g_{\phi t}}{g_{\phi \phi}}~~.
\end{equation}
For the particular metric used,
\begin{equation}
	ds^2 =	-e^{\nu-w}( udz^2+dr^2)
		-r^2 e^{-w} d\phi^2+e^w(cdt-Nd\phi)^2~~,
\label{metric}
\end{equation}
Eq. (\ref{eq:omega}) equates with the answer given, \cite{CT},
\begin{equation}
\omega = \frac{N c e^{w} }{r^2 e ^{-w} - N^2 e ^w}~~.
\end{equation}

The error in this derivation is that it does not describe the tangential velocity of a dust particle moving on a geodesic. This is described by the geodesic equation, containing derivatives of the metric in the connection term. Eq. (\ref{eq:omega}) measures frame-dragging rather than geodesic motion, as we now show.

The metric, Eq. (\ref{metric}), does not depend on $\phi$, so a particle geodesic path conserves the momentum component $p_{\phi}$. Consider a zero-momentum particle $p_{\phi}=0$, and note that 
\begin{eqnarray}
p^{\phi} = g^{\phi \phi} p_{\phi} + g^{\phi t} p_t = g^{\phi t} p_t \\
p^t = g^{t t} p_{t} + g^{t \phi} p_\phi = g^{t t} p_{t}~~.
\end{eqnarray}
The $\phi$ coordinate motion of the particle, or frame-dragging, is then described by
\begin{equation}
\frac{d\phi}{d t} = \frac{p^{\phi}}{p^t} = \frac{g^{\phi t}}{g^{t t}}~~.
\end{equation}
Finally we note that in general, by inversion of $g$, $g^{\phi t}/g^{t t} = -g_{\phi t}/g_{\phi \phi}$. Hence Eq. (\ref{eq:omega}) actually measures the frame-dragging experienced by the zero-monentum test particle. Clearly the frame dragging rate is normally much smaller than the circular geodesic rate for non-relativistic systems, so in general these cannot be equated, as seems to have been done by the authors in \cite{CT}.

Another possible area for error is the relationship between redshift measured to determine rotation curves, and the coordinate velocities. This is straightforward in a flat relativistic system, however the presence of frame-dragging, if significant, could complicate matters.

Although it appears that the treatment of tangential velocity is at fault in this case, the question of observing relativistic effects in low density and velocity stationary systems is still relevant. For a large enough rotating system it is likely that the inside can be shielded from the outer universe to some extent by the bulk of the system and thus provide a region that resembles the background locally, a relativistic effect. Whether this can be achieved without requiring relativistic velocities relative to the background universe is less clear.

\end{document}